\documentclass[12pt]{iopart}
\usepackage{graphicx}
\usepackage{dcolumn}
\usepackage{bm}
\usepackage{amssymb}
\usepackage{xfrac}
\usepackage{float}
\usepackage{iopams}
\usepackage{cite}

\begin{document}
\title[]{On the role of spatial dynamics and topology on network flows}

\author{Serdar \c{C}olak$^1$\footnote{Author to whom any correspondence should be addressed.}, Christian M Schneider$^1$, Pu Wang$^2$, Marta C Gonz\'{a}lez$^{1,3}$}

\address{$^1$Department of Civil and Environmental Engineering, Massachusetts Institute of Technology, Cambridge, Massachusetts 02139, USA}
\address{$^2$School of Traffic and Transportation Engineering, Central South University, Hunan 410083, China}
\address{$^3$Engineering Systems Division, Massachusetts Institute of Technology, Cambridge, Massachusetts 02139, USA}
\ead{serdarc@mit.edu}

\begin{abstract}
Particle flows in spatial networks are susceptible to congestion. In this paper, we analyze the phase transitions of these networks to a state of congested transport and the influence of both topology and spatial dynamics on its emergence. We systematically show that the value of the critical loading rate at which congestion emerges is affected by the addition of spatial dynamics, changing the nature of this transition from a continuous to a discontinuous one. Our numerical results are confirmed by introducing an analytical solvable framework. As a case of study, we explore the implications of our findings in the San Francisco road network where we can locate the roads that originate the congested phase. These roads are the spatially constrained, and not necessarily those with high betweenness as predicted by models without spatial dynamics.
\end{abstract}

\vspace{2pc}

\maketitle

\section{Introduction}

Flow networks are inherently liable to congestion. The ability of these networks to handle demand at reasonable levels is crucial as otherwise a congested phase of transport affects the performance across the entire network. Therefore, it is of prime interest to analyze how and where networks begin to undergo a transition to a congested state and the dynamics of its response. Dissecting flow patterns is essential to address this problem. In this context, flow of data packets in the Internet is well understood, as analyses of their traffic dynamics and phase transitions are abundant \cite{sawatari, echenique, sreenivasan2, bianconi, sreenivasan, mendes_gridlock}. The transition point to congestion is in this case well established through analytical solutions and simulations \cite{guimera1, arenas2, guimera2, zhao, zhou}. To interpret the role of a network backbone for managing flows, optimal paths and minimum spanning trees have been studied  \cite{havlin3, havlin2, havlin1}. The most relevant metric to determine the vulnerability of internet flows is the betweenness centrality for shortest paths, because it determines the critical element generating congestion in the network \cite{freeman, barthelemy}. 

All these previous studies have overlooked the role of space and time in particle flows in the networks; a reasonable assumption in Internet applications. In this case the travel time between nodes is negligible and queues accumulate in the nodes which have limited capacity to process the packets and congestion occurs. In contrast, in transportation networks travel time of vehicles or individuals is crucial. Queues are formed on the links which have limited spatial capacity, and in turn their velocity depends on the density of travelers on each link. In this context, transportation research focuses on capturing traffic flow by making use of the fundamental diagrams that empirically relate flow, density and speed in road segments and utilizing them in macroscopic link models \cite{daganzo_fund, daganzo_fund2}. The cell transmission model \cite{daganzo, daganzo2} and the simple point queue models \cite{kuwahara2, kuwahara, zhang} are well established among such traffic flow models. Alternatively, cellular automata models for vehicular traffic \cite{nagel1, chowdhury} have also been used to mimic traffic flow behavior along with many other discrete stochastic models \cite{nagel2, nagel0}. In none of these cases, the interaction of the spatial dynamics with the network topology have been addressed. Our goal here is to build on the simple model of the Internet by adding temporal and spatial dimensions of particle flow and hence construct a framework to analyze congestion in spatial networks from a network science perspective. The proposed framework can be extended to flows in other kinds of spatial networks \cite{nature_transport, barthelemy_report, christian}, and more importantly could open new avenues of the research on urban road networks  that would go beyond modeling and topological analysis for the statistical physics community\cite{rosvall2005networks, barthelemy2008modeling, youn2008price}.

\section{Models}

\begin{figure}[b!]
\centering
\includegraphics[scale=0.3 , trim = 20mm 00mm 0mm 0mm]{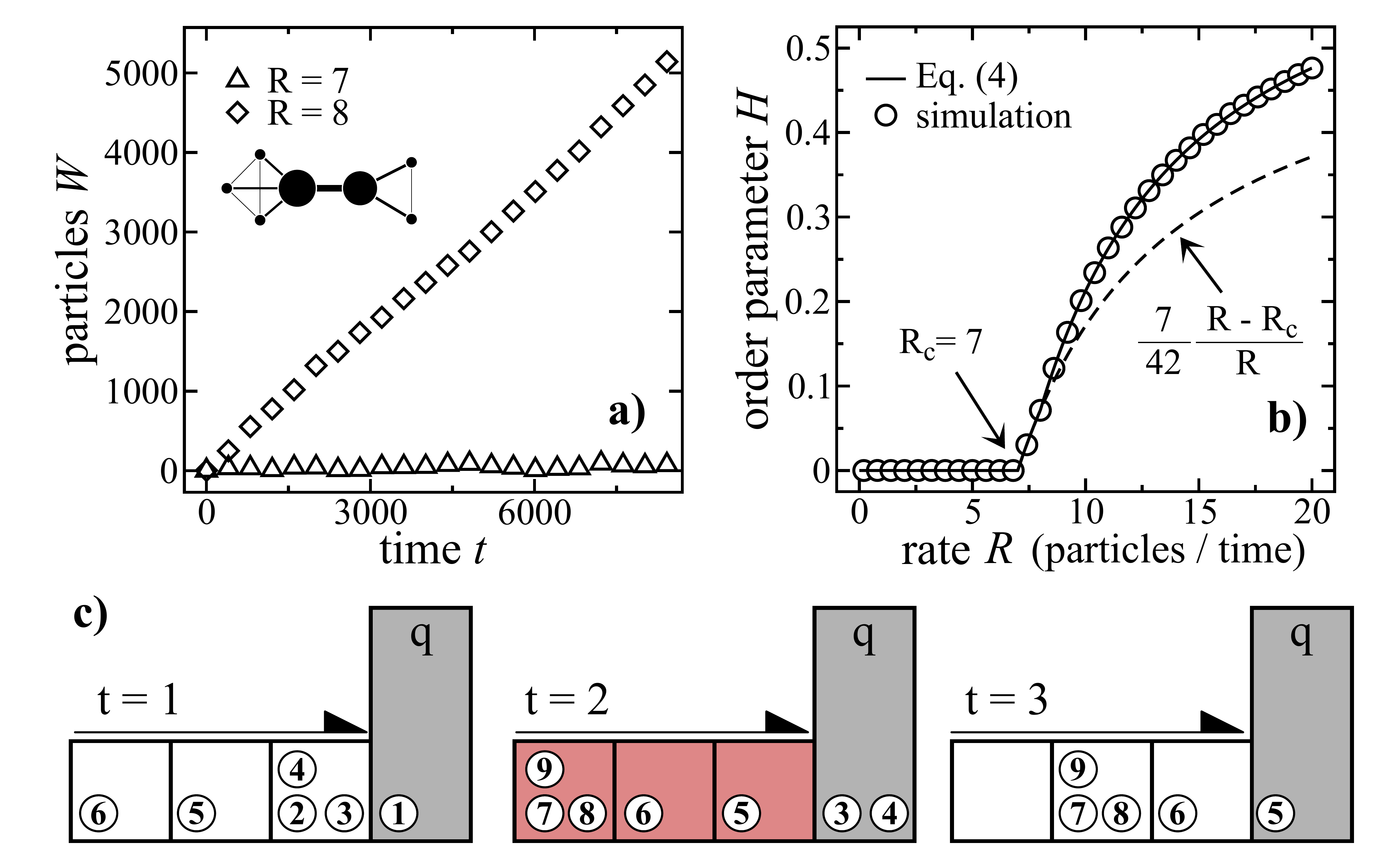}
\caption{\label{fig1} (a) Number of particles $W$ versus time $t$ for the Internet model with different loading rates for the shown simple network with node and edge betweenness values mapped as node size and edge width, respectively. (b) A second order transition at the critical value $R_{c}=7$ for this network. (c) A schematic representation of the SPQM for a link with $C=2$, $\tau=3$ and $V=7$. At $t=2$, the link reaches its volume capacity and no incoming particle is allowed entering the link.}
\end{figure}

We begin by recalling the scheme in \cite{guimera1, guimera2, sreenivasan, sreenivasan2, zhou, zhao, arenas2, bianconi, sawatari, echenique} and refer to it as the internet model (IM): the network is loaded with $R$ identical particles at each timestep $t$ with randomly assigned origins and destinations. A fixed shortest path routing table guides particles towards their destination. Nodes can transmit as many particles per timestep as their \textit{outflow capacity}, $C$, and travel between two nodes takes a unit timestep. Queues of particles form at the nodes, and they can grow infinitely large. Particles are exempt from joining the queue at their destination and are removed from the system upon arrival. The network response is measured by the order parameter $H$ \cite{guimera2}: \begin{equation} H\left(R\right) = \lim \limits_{t \to T} \frac{\left\langle \Delta W \right\rangle}{R \Delta t}, \end{equation} where $W$ denotes the number of particles in the system, $\left\langle \Delta W \right\rangle$ is the average change in the number of particles still in the system after a timestep,  $\Delta t$ is the unit timestep and $T$ is the length of the simulation. Figure \ref{fig1} (a) and (b) depict the IM. For low values of $R$, the network reaches a rate of particle arrival equal to the loading rate. $W$ remains constant and consequently $H=0$. Conversely, if $R$ exceeds a certain threshold $R_{c}$, a linear increase in $W$ with a slope of $H$ is observed due to excessive queueing.  This behavior maps a second-order phase transition to congestion.

Spatial networks, being embedded in two-dimensional space, give rise to three heterogeneities that need to be captured. First is non-uniform travel times. This differs from the Internet where data packets hop from one node to the next in a single timestep. Second, these networks carry flows along the links. A particle in a spatial networks has a specific position on the link it is traveling on. The third source of heterogeneity is a consequence of this: particles occupy physical space and gradually fill the segment. Among the various models of traffic flow in the transportation literature aimed to address these issues, the point-queue model (PQM) used in \cite{zhang, kuwahara, kuwahara2, drissi} is an adaptation of the IM that shifts flow from nodes to links, incorporating the non-uniform travel time distribution and thereby making the flow analysis very similar to that of the internet. Particles traverse the link freely by hopping through $\tau$ slots of unit travel time to join a queue at the end of the link from which they will be discharged at the outflow capacity. The total travel time consists of the free travel time and the delay, namely, the timespan between the particle entering the queue and exiting it. The spatial point-queue model (SPQM) incorporates a single additional constraint to the PQM: every link has an upper limit for the number of particles it can hold at once. We will refer to this value as the \textit{volume capacity} of the link, $V$. Links cannot accept any new particles when they reach their volume capacity, as illustrated in Figure \ref{fig1} (c). This additional constraint has a crucial effect on the nature of the network response, as links at volume capacity clog upstream links and cause them to succumb to congestion as well. This spreading of congestion occurs at rates that depend on the loading rate and the network topology along with specific link properties. The rate of particles unable to travel determines the speed with which the congestion spills, which makes the spreading process non-binary unlike traditional spreading models in the literature. Although the SPQM share some aspects with several directed percolation models, the movement of non-identical particles along predetermined spatial shortest paths with non-binary spreading is uncommon and therefore relatively unstudied  \cite{havlindiffusion, hinrichsen}. 

\section{Results and discussion}

The critical loading rate $R_{c}^{\text{IM}}$ has been shown \cite{guimera1} to be equal to $R_{c}^{\text{IM}}=N(N-1)(C_{\text{max}}/B_{\text{max}}^{\text{N}})$ where $N$ is the network size measured by the number of nodes, $B_{\text{max}}^{\text{N}}$ is the maximum node betweenness and $C_{\text{max}}$ is the outflow capacity of this node. This relation arises from the fact that inflow to a node is proportional to its betweenness centrality \cite{goh}. At $R_{c}$, the inflow is equal to the outflow at the node with the minimum $C_{\text{max}}/B_{\text{max}}^{\text{N}}$ value. For the PQM, we adjust this equation by replacing the node betweenness by an edge betweenness value, $R_{c}^{\text{PQM}}= N(N-1)(C_{\text{max}}/B_{\text{max}}^{\text{E*}})$, where $B_{\text{max}}^{\text{E*}}$ refers to the maximum \textit{modified betweenness}. This modification is necessary since particles do not join the queues in the final links, which should be omitted from the edge betweenness calculations. 

\begin{figure}[b]
\centering
\includegraphics[scale=0.45, trim = 0mm 0mm 0mm 0mm]{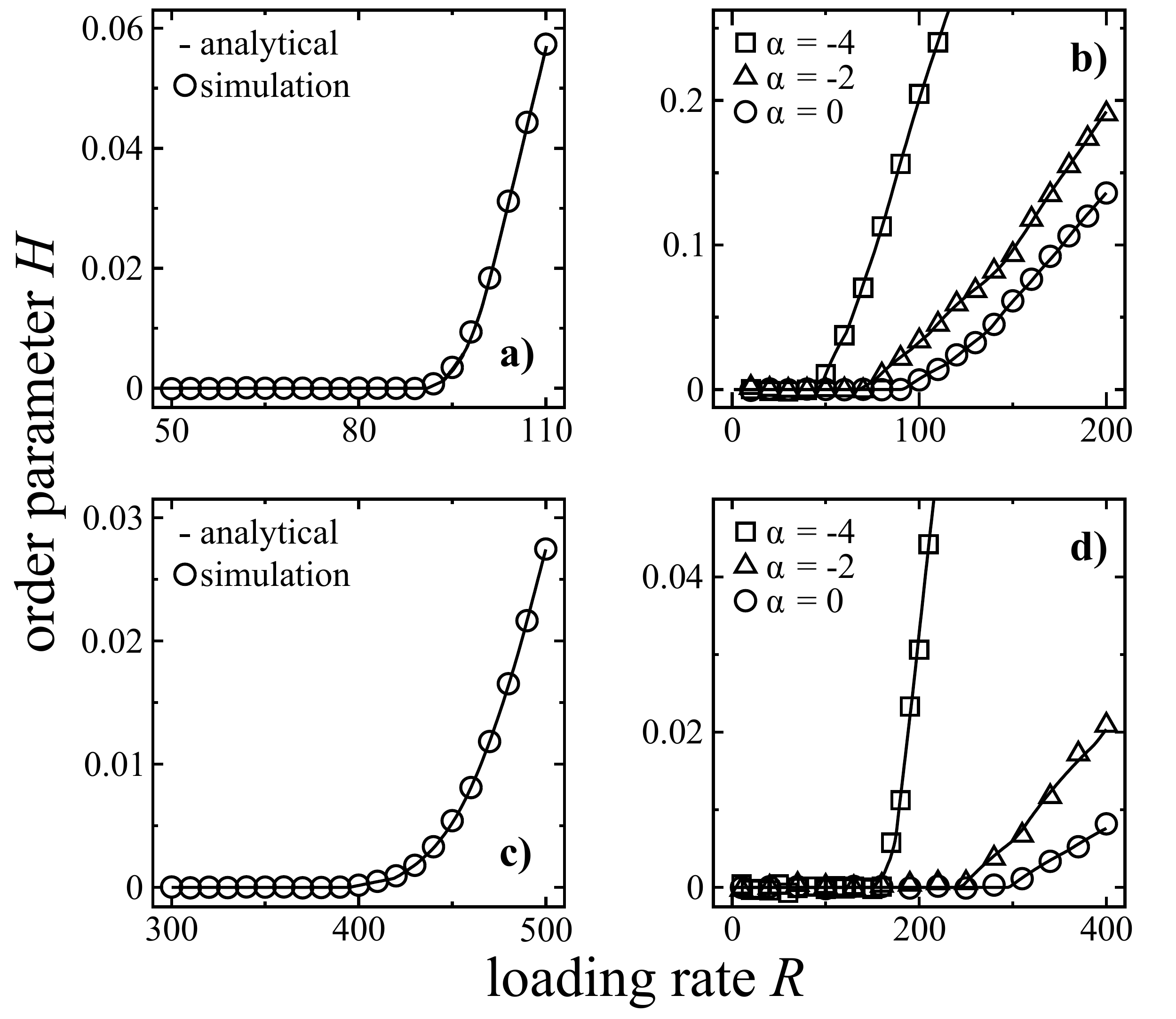}
\caption{\label{fig2} Transitions of a non-periodic lattice as estimated by simulations and by analytical solutions for (a) the IM and (c) the PQM. Transitions of a completely rewired network with $\rho=1$ for (b) the IM and (d) the PQM for varying tendencies $\alpha$ to have long distance links. Analytical solutions are consistent with simulations.}
\end{figure}

Next, we introduce a framework to analytically calculate the entire transition curve to congestion. For $R>R_{c}$, particle inflow at certain elements will be larger than the outflow. We define $R_{c}^{i}$ as the critical loading rate specific to element $i$. For large $R$, the outflow of congested links are maximized to capacity, which in consequence affects the inflow to the links downstream. To account for this we define the \textit{delay factor}, $D_{i}(R)$, referring to the fraction of paths through $i$ that are not suffering from delay as, \begin{equation} \label{a1} D_{i}(R) = \mathcal{H} ( R_{c}^{i}-R)+\frac{C_{i}}{I_{i}(R)} \mathcal{H} ( R-R_{c}^{i}), \end{equation} where $\mathcal{H}(x)$ is the Heaviside step function and $C_{i}$ and $I_{i}(R)$ are the outflow capacity and the inflow of element $i$ for loading rate $R$. $D_{i}(R)=1$ suggests no congestion for element $i$, whereas lower values indicate levels of congestion. Using this definition, the inflow $I_{i}(R)$ at a specific loading rate $R$ can be quantified as, \begin{equation} \label{a2} I_{i}(R) = \frac{\sum_{k \in \Gamma(i)} \prod_{j \in k} R  D_{j}(R)}{N(N-1)} \end{equation} \begin{equation} \label{a3} H(R) = \sum_{i \in \mathcal{N}} (I_{i}(R)-C_{i}) \mathcal{H}(I_{i}(R)-C_{i}), \end{equation} where $\Gamma(i)$ is the set of paths passing through element $i$. Eq.(\ref{a3}) accounts for all the delay factors of the elements upstream of element $i$ by going through the shortest paths. Eq. (\ref{a1}) and (\ref{a2}) form a set of coupled equations that can be solved to obtain the inflows for every element. $H(R)$ is obtained by summing all positive values of  $I_{i}(R)-C_{i}$. Figure \ref{fig1} (b) also reveals the exact solution for the simple network around the critical point with the dashed curve $H=(7/42)(1-R_{c}/R)$. In order to test different network topologies and examine the effect of space on criticality, we use a non-periodic lattice as a substrate and rewire each edge $(i,j)$ with probability $\rho$ to a new destination $j^{*}$ chosen with probability proportional to $d(i,j^{*})^{\alpha}$ where $d(i,j)$ denotes the Euclidean distance \cite{watts, kleinberg}. Figure \ref{fig2} reveals that the simulations and analytical results perfectly coincide for transitions in both a two dimensional non-periodic lattice of $N=1225$ and $C=4$, and its completely rewired instance. Critical rates decrease for smaller $\rho$ as newly introduced shortcuts have higher betweenness values. As $\rho$ increases, the value of $\alpha$ and its effect on network topology becomes more pronounced: rewired links in networks with lower $\alpha$ values are more localized and therefore maximum betweenness values are higher in these networks. Consequently the increase in $H$ is sharper for lower $\alpha$. Figure \ref{fig2} (b) and (d) show sharper transitions and lower $R_{c}$ for decreasing $\alpha$ values for network instances with $\rho=1$.

\begin{figure}
\centering
\includegraphics[scale=0.3, trim = 20mm 10mm 0mm 20mm]{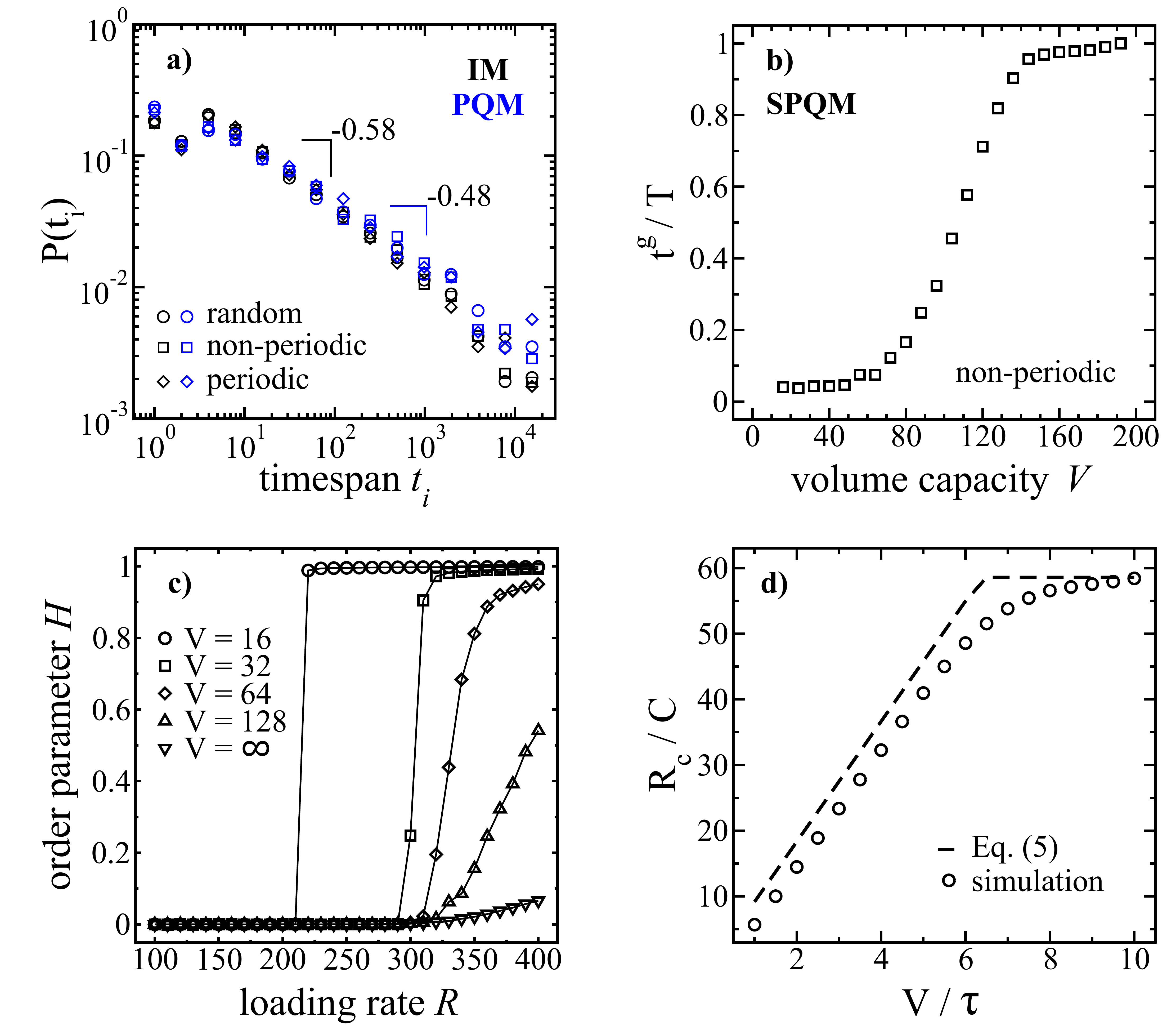}
\caption{(a) Distributions of the timespans through which the congested element operates at its outflow capacity at $R_{c}$ for the IM and PQM, following a power law. (b) The average number of timesteps normalized by $T$ before a gridlock is observed, versus the volume capacity.  (c) Change in the nature of the transition to congestion of a non-periodic lattice of size $N=625$ with specified levels of volume capacities. (d) The critical loading rate $R_{c}$ for varying $V/\tau$ values. $R_{c}$ settles to the PQM-limit as $V/\tau$ increases. }
\label{fig3}
\end{figure}

At $R_{c}$, the link that triggers congestion, also referred to as the critical element, is expected to fluctuate between free flow and congested phases. Figure \ref{fig3} (a) exhibits the frequency distribution of the timespans at which this most critical element operates at its outflow capacity, as an indicator of the temporality of the phase transition. Results show that these timespans follow a power law with exponents of $-0.58\pm0.04$ for the IM and $-0.48\pm0.04$ for the PQM, independent of the network topology. SPQM exhibits a different behavior. For low volume capacities links tend to fill up, causing links upstream to fill as well. Fluctuations cause a \textit{gridlock}, a condition where all elements of a cycle are completely filled and hence flows stop. In case of a gridlock, the order parameter increases very sharply. In Figure \ref{fig3} (b), for varying volume capacities we measure the average number of timesteps it takes for a gridlock to occur, $t^{g}$, normalized by the length of the simulation $T$. It can be observed that for lower $V$, gridlocks are observed relatively quickly. For large volume capacities, PQM and SPQM have the same $R_{c}$, which we will refer to as the PQM-limit. In either model, at steady state, queues are not expected to be necessarily empty but rather steady in their size. If the volume capacity of a link is smaller than this steady state queue size, particles will be blocked in the upstream link which consequently may suffer from a decrease in its outflow due to this clogging effect. This suggests a lower critical point for a network with active volume capacity constraints, correspondingly $R_{c}^{\text{PQM}} \geq R_{c}^{\text{SPQM}}$. Therefore, to realize the congestion-free transport to the fullest, the PQM-limit should be aimed. Figure \ref{fig3} (c) illustrates the response of a non-periodic lattice with $N=625$ for decreasing levels of volume capacity and reveals that both the critical loading rate and the nature of the transition is affected by $V$. Figure \ref{fig3} (d) shows the effect of the volume levels on $R_{c}$. In point queue models, the number of particles on a link is the sum of those in the queue, which was shown to be proportional to the modified betweenness, and those that are traveling. The expected inflow to a single slot of travel on a link is proportional to the actual betweenness of that link, hence the expected number of particles traveling is $\tau B^{E}$. Therefore for small $V$ and strictly deterministic inflows, a link reaches its maximum volume when $V=R_{c}\left(\tau B^{E}+B^{E*}\right)_{\text{max}}/N(N-1)$.  The inflow to the link is proportional to the actual routing betweenness so the critical point can be expressed as, \begin{equation} \label{eq5} R_{c} = N(N-1)\min \left( \frac{C}{B^{E*}_{\text{max}}}, \frac{V}{(\tau B^{E}+B^{E*})_{\text{max}}} \right), \end{equation}  where $B^{E}$ is the edge betweenness. Consequently a linear increase in $R_{c}$ is observed up to the point where $V=C(\tau B^{E}+B^{E*})_{\text{max}}/B^{E*}_{\text{max}}$. However, the stochastic nature of the model, along with the first-order nature of its transition, causes fluctuations that force $R_{c}/C$ below this analytical bound as it converges to the PQ-limit.

\begin{figure}
\centering
\includegraphics[scale=0.35, trim = 0mm 10mm 0mm 0mm]{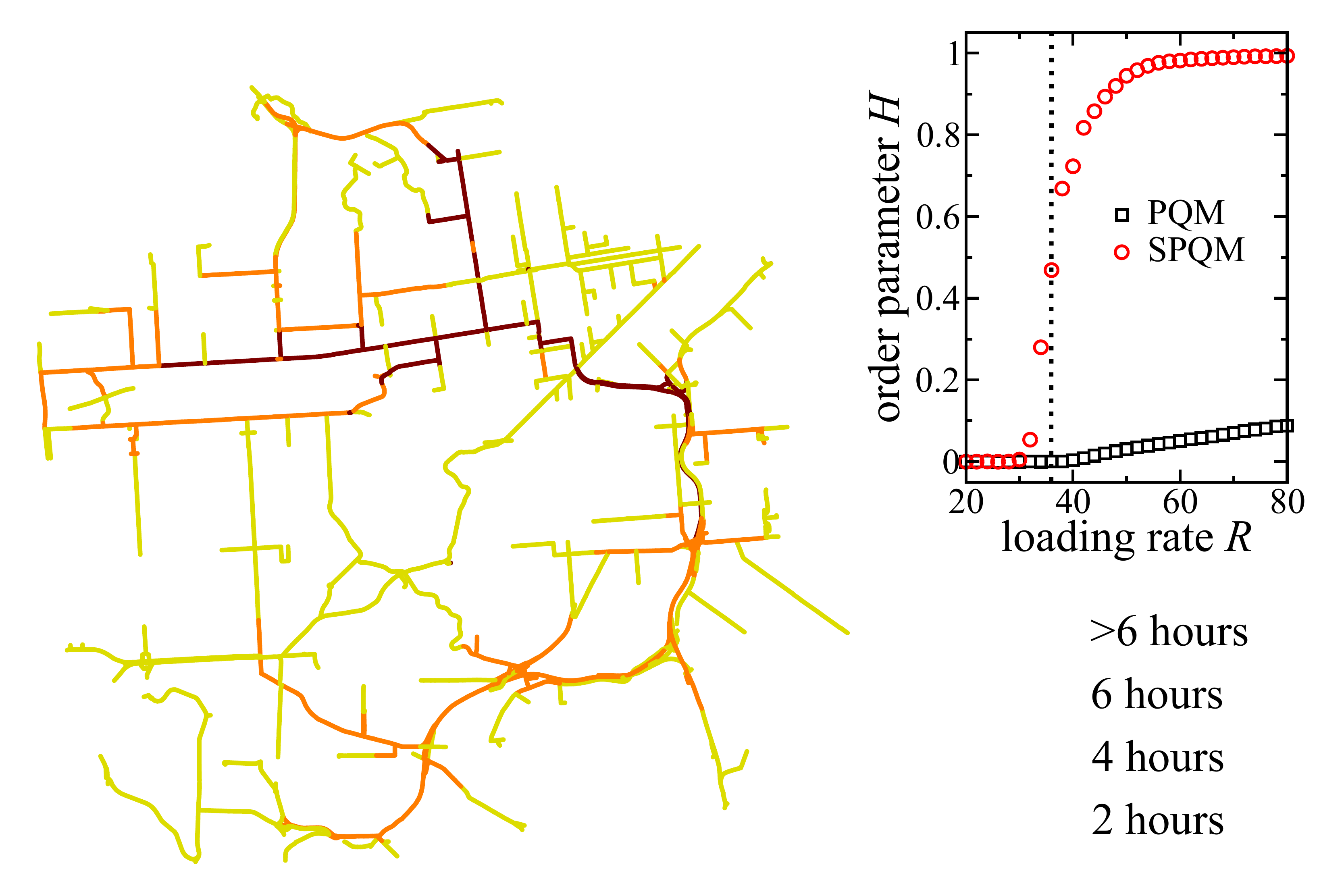}
\caption{A congestion map of the San Francisco road network for $R=36$ vehicles per timestep. Colors represent the times within which the road segments become congested as a consequence of the spillover. The black circle denotes the origin of congestion. (inset) PQM and SPQM transitions for the San Francisco road network, where $R_{c}^{\text{PQM}}=40$ and $R_{c}^{\text{SPQM}}=30$.}
\label{fig4}
\end{figure}

Figure \ref{fig4} depicts the transitions for the PQM and the SPQM for the San Francisco road network with $N=1152$ and an average degree of $3.2$. The network is discretized by unit travel times of 10 seconds. Outflow capacity of a road segment is obtained by using the speed limit and the number of lanes. Volume capacities are estimated for every road segment assuming that the volume capacity is reached when speed drops to half of the speed limit. Under these assumptions, the PQM-limit is not reached as $R_{c}^{\text{PQM}}=40$ (14400 vehicles/h) and $R_{c}^{\text{SPQM}}=30$ (10800 vehicles/h).  To capture network response in the SPQM, segment volumes are recorded at different time steps of the simulation $t=720$ (2h), $t=1440$ (4h) and $t=2160$ (6h) for $R=36$, slightly above the critical load. Figure \ref{fig4} illustrates the network response by mapping road segments that have reached 80 percent of their volume capacities in the given time periods by color. Congestion originates from an artery leading to the downtown area and anisotropically spreads to other regions. After 6 hours, most network elements are suffering from congestion.

\section{Conclusion}

Our findings suggest that in the SPQM, the critical road segment cannot accept incoming vehicles as it saturates to its volume capacity before it reaches its outflow capacity. This outcome can be traced back to the inherence of congestion in downtown areas: cities with high population densities have concentrated spatial demand distributions, which result in the inadequacy of urban space to accommodate such concentrated flows. This work is a step further on a systems analysis applied to congestion in roads. In further studies, population and facility distributions can be modeled. In the applications domain an open question is to know how the studied transition to congestion is influenced by introducing empirical origin-destination matrices that represent the real population's travel demand \cite{puwang, marta_radiation} and how network topology contributes towards reducing- or maybe aggravating- congestion. While in the domain of phase transitions, analyzing the set of critical exponents using directed percolation as a benchmark case remains as an intriguing and open question \cite{havlindiffusion, hinrichsen}.

\section{Acknowledgements}

This work was funded by New England UTC Year 24 grant, awards from NEC Corporation Fund, the Solomon Buchsbaum Research Fund, BMW, UPS and the National Natural Science Foundation of China (No. 51208520). The authors would like to thank Zolt\'{a}n Toroczkai and Ruben Juanes for useful comments and technical support.

\section*{References}

\bibliography{cong}{}

\providecommand{\newblock}{}
\begin{thebibliography}{10}
\expandafter\ifx\csname url\endcsname\relax
  \def\url#1{{\tt #1}}\fi
\expandafter\ifx\csname urlprefix\endcsname\relax\def\urlprefix{URL }\fi
\providecommand{\eprint}[2][]{\url{#2}}

\bibitem{sawatari}
Ohira T and Sawatari R 1998 {\em Phys. Rev. E\/} {\bf 58}(1) 193--195

\bibitem{echenique}
Echenique P, G{\'o}mez-Garde{\~n}es J and Moreno Y 2005 {\em Europhys. Lett.\/}
  {\bf 71} 325

\bibitem{sreenivasan2}
Sreenivasan S, Cohen R, L{\'o}pez E, Toroczkai Z and Stanley H~E 2007 {\em
  Phys. Rev. E\/} {\bf 75} 036105

\bibitem{bianconi}
De~Martino D, Dall’Asta L, Bianconi G and Marsili M 2009 {\em Phys. Rev. E\/}
  {\bf 79} 015101

\bibitem{sreenivasan}
Asztalos A, Sreenivasan S, Szymanski B~K and Korniss G 2012 {\em Eur. Phys. J.
  B\/} {\bf 85} 1--10

\bibitem{mendes_gridlock}
Mendes G, Da~Silva L and Herrmann H 2012 {\em Phys. A\/} {\bf 391} 362--370

\bibitem{guimera1}
Arenas A, D{\'\i}az-Guilera A and Guimer{\`a} R 2001 {\em Phys. Rev. Lett.\/}
  {\bf 86} 3196--3199

\bibitem{arenas2}
Guimer\`{a} R, Arenas A, D\'{\i}az-Guilera A and Giralt F 2002 {\em Phys. Rev.
  E\/} {\bf 66} 026704+

\bibitem{guimera2}
Guimer{\`a} R, D{\'\i}az-Guilera A, Vega-Redondo F, Cabrales A and Arenas A
  2002 {\em Phys. Rev. Lett.\/} {\bf 89} 248701

\bibitem{zhao}
Zhao L, Lai Y~C, Park K and Ye N 2005 {\em Phys. Rev. E\/} {\bf 71}(2) 026125

\bibitem{zhou}
Yan G, Zhou T, Hu B, Fu Z~Q and Wang B~H 2006 {\em Phys. Rev. E\/} {\bf 73}
  046108+

\bibitem{havlin3}
Tomer E, Safonov L, Madar N and Havlin S 2002 {\em Phys. Rev. E\/} {\bf 65}
  065101

\bibitem{havlin2}
Wu Z, Braunstein L~A, Havlin S and Stanley H~E 2006 {\em Phys. Rev. Lett.\/}
  {\bf 96} 148702+

\bibitem{havlin1}
Carmi S, Wu Z, Havlin S and Stanley H~E 2008 {\em Europhys. Lett.\/} {\bf 84}
  28005+

\bibitem{freeman}
Freeman L~C 1977 {\em Sociometry\/} {\bf 40} 35--41

\bibitem{barthelemy}
Barth{\'e}lemy M 2004 {\em Eur. Phys. J. B\/} {\bf 38} 163--168

\bibitem{daganzo_fund}
Geroliminis N and Daganzo C~F 2008 {\em Transp. Res. B\/} {\bf 42} 759--770

\bibitem{daganzo_fund2}
Daganzo C~F and Geroliminis N 2008 {\em Transp. Res. B\/} {\bf 42} 771--781

\bibitem{daganzo}
Daganzo C~F 1994 {\em Transp. Res. B\/} {\bf 28}(4) 269--287

\bibitem{daganzo2}
Daganzo C~F 1995 {\em Transp. Res. B\/} {\bf 29} 79--93

\bibitem{kuwahara2}
Kuwahara M and Akamatsu T 1997 {\em Transp. Res. B\/} {\bf 31} 1--10

\bibitem{kuwahara}
Kuwahara M and Akamatsu T 2001 {\em Transp. Res. B\/} {\bf 35} 461--479

\bibitem{zhang}
Nie X and Zhang H~M 2005 {\em Networks and Spatial Economics\/} {\bf 5} 89--115

\bibitem{nagel1}
Nagel K and Schreckenberg M 1992 {\em J. Phys. I\/} {\bf 2} 2221--2229

\bibitem{chowdhury}
Chowdhury D, Santen L and Schadschneider A 2000 {\em Phys. Rep.\/} {\bf 329}
  199--329

\bibitem{nagel2}
Schreckenberg M, Schadschneider A, Nagel K and Ito N 1995 {\em Phys. Rev. E\/}
  {\bf 51} 2939

\bibitem{nagel0}
Nagel K 1996 {\em Phys. Rev. E\/} {\bf 53} 4655

\bibitem{nature_transport}
Banavar J~R, Maritan A and Rinaldo A 1999 {\em Nature\/} {\bf 399} 130--132

\bibitem{barthelemy_report}
Barth{\'e}lemy M 2011 {\em Phys. Rep.\/} {\bf 499} 1--101

\bibitem{christian}
Mamede G~L, Ara{\'u}jo N~A, Schneider C~M, de~Ara{\'u}jo J~C and Herrmann H~J
  2012 {\em Proc. Nat. Acad. USA\/} {\bf 109} 7191--7195

\bibitem{rosvall2005networks}
Rosvall M, Trusina A, Minnhagen P and Sneppen K 2005 {\em Phys. Rev. Lett.\/}
  {\bf 94} 028701

\bibitem{barthelemy2008modeling}
Barth{\'e}lemy M and Flammini A 2008 {\em Phys. Rev. Lett.\/} {\bf 100} 138702

\bibitem{youn2008price}
Youn H, Gastner M~T and Jeong H 2008 {\em Phys. Rev. Lett.\/} {\bf 101} 128701

\bibitem{drissi}
Drissi-Ka{\"\i}touni O and Hameda-Benchekroun A 1992 {\em Transp. Sci.\/} {\bf
  26} 119--128

\bibitem{havlindiffusion}
Havlin S and Ben-Avraham D 1987 {\em Adv. Phys.\/} {\bf 36} 695--798

\bibitem{hinrichsen}
Hinrichsen H 2000 {\em Adv. Phys.\/} {\bf 49} 815--958

\bibitem{goh}
Goh K~I, Kahng B and Kim D 2001 {\em Phys. Rev. Lett.\/} {\bf 87}(27) 278701

\bibitem{watts}
Watts D~J and Strogatz S~H 1998 {\em Nature\/} {\bf 393} 440--442

\bibitem{kleinberg}
Kleinberg J~M 2000 {\em Nature\/} {\bf 406} 845--845

\bibitem{puwang}
Wang P, Hunter T, Bayen A~M, Schechtner K and Gonz{\'a}lez M~C 2012 {\em
  Scientific Reports\/} {\bf 2} 1001

\bibitem{marta_radiation}
Simini F, Gonz{\'a}lez M~C, Maritan A and Barab{\'a}si A~L 2012 {\em Nature\/}
  {\bf 484} 96--100

\end{thebibliography}
\bibliographystyle{cong}

\end{document}